# Spin-valley-controlled photonic topological insulator

Haoran Xue[1,2,#], Fei Gao[1,5,#,*], Yang Yu[3], Yidong Chong[2,4], Gennady Shvets[3,*], Baile Zhang[2,4,*]

[1]State Key Laboratory of Modern Optical Instrumentation, and College of Information Science and Electronic Engineering, Zhejiang University, Hangzhou 310027, China.

[2]Division of Physics and Applied Physics, School of Physical and Mathematical Sciences, Nanyang Technological University, Singapore 637371, Singapore.

[3]School of Applied and Engineering Physics, Cornell University, Ithaca NY 14853.

[4]Centre for Disruptive Photonic Technologies, Nanyang Technological University, Singapore 637371, Singapore.

[5]Key Lab. of Advanced Micro/Nano Electronic Devices & Smart Systems of Zhejiang, The Electromagnetics Academy at Zhejiang University, Zhejiang University, Hangzhou 310027, China.

[#]Authors contributed equally.

*Author to whom correspondence should be addressed; E-mail: gaofeizju@zju.edu.cn (F. Gao) gshvets@cornell.edu (G. Shvets); blzhang@ntu.edu.sg (B. Zhang)

**The discovery of photonic topological insulators (PTIs)[1-3] has opened the door to fundamentally new topological states of light. Current time-reversal-invariant PTIs emulate either the quantum spin Hall (QSH) effect[4-11] or the quantum valley Hall (QVH) effect[12-17] in condensed-matter systems, in order to achieve topological transport of photons whose propagation is predetermined by either 'photonic pseudospin' (abbreviated as 'spin') or valley. Here we demonstrate a new class of PTIs, whose topological phase is not determined solely by spin or valley, but is controlled by the competition between their induced gauge fields. Such a competition is enabled by tuning the strengths of spin-orbit coupling (SOC) and inversion-symmetry breaking in a single PTI[18,19]. An unprecedented topological transition between QSH and QVH phases that is hard to achieve in condensed-matter systems is demonstrated. Our study merges the emerging fields of spintronics and valleytronics in the same photonic platform, and offers novel PTIs with reconfigurable topological phases.**

In electronic systems, spin and valley are two kinds of binary degrees of freedom (DOFs) that have played a key role in the past decade in designing and constructing novel topological phases. Take graphene as a typical example. In 2005, it has been shown that by introducing SOC, the well-known Dirac cones of graphene can open a bandgap[19] (see Figs. 1a-b; details to be discussed later) that is topologically nontrivial since the integration of Berry curvature over the whole Brillouin zone for a specific spin state, known as spin Chern number, is nonzero; this is now termed as the QSH effect, or widely referred to as the 2D topological insulator[20,21]. In parallel, it has been known since 2007 that breaking the inversion symmetry with staggered sub-lattice potential in graphene can also produce a bandgap[22,23] (see Fig. 1c; details to be discussed later) near the Dirac points. Although the full Chern number is zero, this bandgap is topologically nontrivial when projected to a specific valley. The topological valley transport has also been recently demonstrated in biased bilayer graphene[24,25], as nowadays called the QVH effect, a fundamentally new topological transport in the emerging field of valleytronics[26,27].

To further understand the underlying physics, we can resort to an effective Hamiltonian around K and K' valleys[13,19], which describes a Dirac cone with two different perturbations:

$$\mathcal{H}(\delta k) = \mathcal{H}_0 + \mathcal{H}_{SOC} + \mathcal{H}_P, \tag{1}$$

where $\mathcal{H}_0 = v_D(\delta k_x \hat{\tau}_z \hat{s}_0 \hat{\sigma}_x + \delta k_y \hat{\tau}_0 \hat{s}_0 \hat{\sigma}_y)$ represents gapless Dirac Hamiltonian, and $\mathcal{H}_{SOC} = \Delta_{SOC} \hat{\tau}_z \hat{s}_z \hat{\sigma}_z$ and $\mathcal{H}_P = \Delta_P \hat{\tau}_0 \hat{s}_0 \hat{\sigma}_z$ describe the perturbation of SOC and inversion-symmetry breaking, respectively. Here $\hat{\tau}_{x,y,z}$, $\hat{s}_{x,y,z}$, $\hat{\sigma}_{x,y,z}$ are Pauli matrices acting on valley, spin and orbital subspaces, and $\hat{\tau}_0$ and $\hat{s}_0$ are unity matrices acting on valley and spin subspaces. $v_D$ is the group velocity near Dirac point. $\Delta_{SOC}$ and $\Delta_P$ are the effective masses induced by SOC and inversion-symmetry breaking, respectively. Interestingly, this Hamiltonian takes the same form as the one of silicene without Rashba terms[28], which means a photonic realization of equation (1) can also lead to rich topological phases as predicted in silicene[28].

In absence of both SOC and inversion-symmetry breaking, the gapless Hamiltonian $\mathcal{H}_0$ will form Dirac cones at K and K' valleys, as shown in Fig. 1a. When only SOC is included, the bandgap opened in Fig. 1b is fully determined by SOC, and thus has the bandgap size $2|\Delta_{SOC}|$. On the other hand, when only inversion-symmetry breaking is present, the bandgap in Fig. 1c carries a bandgap size $2|\Delta_P|$, as determined by the strength of inversion-symmetry breaking. Note that in both cases, the bandgap size applies to both spin up and spin down states at both K and K' valleys. However, when SOC and inversion-symmetry breaking are involved simultaneously, according to equation (1), the bandgaps for the two spin states at two valleys are proportional to $|s\Delta_{SOC} + v\Delta_P|$, where $s = \pm 1$ stand for up/down spin states and $v = \pm 1$ represent K and K' valleys, as shown in Fig. 1d. That denotes the spin-valley locking of bulk bands. Furthermore, the present topological phase is determined not solely by $|\Delta_{SOC}|$ or $|\Delta_P|$, but instead by their relative size. One can calculate the valley-projected spin Chern numbers as $C_{s,v} = \frac{1}{2}\text{sgn}(s\Delta_{SOC} + v\Delta_P)$ (see details in Supplementary Information). The results show that when $|\Delta_{SOC}| > |\Delta_P|$, a nonzero value of full spin Chern number $C_s = C_{s,K} + C_{s,K'} = \pm \text{sgn}(\Delta_{SOC})$ emerges, giving rise to the QSH phase; in this case, the bandgap is topologically equivalent to that in Fig. 1b. However, when $|\Delta_{SOC}| < |\Delta_P|$, the two valley-projected spin Chern numbers $C_{s,K}$ and $C_{s,K'}$ have opposite signs, and thus their summation leads to the zero value of $C_s$, exhibiting the QVH phase; in this case, the bandgap has topological properties similar to Fig. 1c. By tuning the relative size of $|\Delta_{SOC}|$ and $|\Delta_P|$, it is possible for the bandgap to undergo a topological transition between the QSH and QVH phases. Such a topological transition connects the two emerging fields of spintronics and valleytronics, which is extremely difficult, if not impossible, in condensed-matter systems.

Both the QSH phase in Fig. 1b and the QVH phase in Fig. 1c have found their counterparts in PTI realizations. In a PTI exhibiting QSH phase, the photonic spin can be constructed by in-phase and out-of-phase superposition of transverse-electric (TE) and transverse-magnetic (TM) modes[4]. The SOC can be realized by either constructing bianisotropic metamaterials[5,29] or by breaking the out-of-plane mirror symmetry[6,7]. In a PTI possessing QVH phase, the inversion-symmetry breaking can be realized with a staggered sublattice potential[16,17], or with a triangle-shaped scatter in a unit cell[12-15]. However, a spin-valley locked PTI with a potential topological transition between the QSH and QVH phases, similar to that in Fig. 1d, still remains at the theoretical level[18,19].

Here we demonstrate a new class of PTIs in presence of both SOC and inversion-symmetry breaking. The topological phase in such a PTI is determined by the competition between the gauge fields induced by spin and valley, rather than solely spin or valley. The unit cell of our design is shown in Fig. 2a (see Supplementary Information for the detailed design procedure). It is a triangular lattice of metallic tripods arranged in a parallel-plate waveguide. Each tripod touches the bottom plate of the waveguide, while leaving a gap $g$ from the top plate. This is for the purpose of introducing out-of-plane mirror symmetry breaking that brings in SOC. The presence of SOC manifests the fundamental difference between the current design and the previous valley photonic-crystal demonstration[14-17]. Note that the strength of SOC can be tuned by altering the gap size $g$. The strength of inversion-symmetry breaking can be adjusted by rotating the orientation angle $\theta$ of the tripod. Sweeping $\theta$ and $g$, we can obtain the phase diagram as shown in Fig. 2b. The point of $(\theta, g)=(30°, 2.2$ mm$)$, denoted as 'C' in Fig. 2b, is taken as a typical example to show the spin-valley-locked band structure, as plotted in Fig. 2c. Here, two spin states can be identified through the phase difference between $E_z$ and $H_z$ fields; the simulated phase differences at K point for the four bands labelled as '1' to '4' in Fig. 2c are shown in Fig. 2d. In the phase diagram of Fig. 2b, there is a boundary of vanishing bandgap. It can be shown that the system undergoes a topological transition across this boundary (see detailed calculations of Berry curvature in Supplementary Information). On the left side of the boundary, $C_{\uparrow,v} = 1/2$ and $C_{\downarrow,v} = -1/2$, which correspond to the QSH phase; yet on the right side of the boundary, $C_{\uparrow,v} = \mp 1/2$ and $C_{\downarrow,v} = \mp 1/2$, which match the QVH phase.

We firstly demonstrate the phenomenon of valley-dependent spin splitting in the bulk of PTI, which was predicted in Ref. 18 but still remains unrealized. We take ($\theta$, $g$)=(30°, 2.2 mm), i.e. point C in Fig. 2b, for the demonstration. Fig. 3a shows the simulated $|E_z|$ fields excited by a linear dipole at 6.35GHz, where spin up (down) states can only be excited at K (K') valleys (see Fig. 2c). Due to the trigonal warping distortion[18,30,31] (see Supplementary Information), the launched Bloch waves split into two beams propagating along ΓK and ΓK' directions respectively. Together with the phase difference plot in Fig. 3c, spin splitting is confirmed by the valley-dependent spin flows. To show this phenomenon in experiment, we constructed a lattice consisting of 19×15 tripods; this lattice was surrounded by absorbing materials. We experimentally scanned $E_z$ field along the white line in Fig. 3a, and observed two spit beams in a range of frequencies as shown in Fig. 3b. To confirm these two beams are spin polarized, we measure the phase difference between $E_z$ and $H_z$ at point A and B in Fig. 3c. The measured phase difference at point A (B) is around $\pi$ (0) as shown in Fig. 3d, being consistent with simulation in Figure 2c.

In the following, we demonstrate the topological phase transition. Although hard to measure directly, the bulk topology can manifest itself at the edges through bulk edge correspondence[31]. We firstly take $(\theta, g) = (5°, 2.2 \text{ mm})$ (denoted as point 'A' in Fig. 2b) to demonstrate the QSH phase. We build a domain wall between two PTI domains with opposite settings, as shown in the inset of Fig. 4a. In the upper domain, the valley-projected spin Chern numbers are $C_{\uparrow/\downarrow,K}=\mp 1/2$ and $C_{\uparrow/\downarrow,K'}=\mp 1/2$, while in the lower domain, they take opposite values: $C_{\uparrow/\downarrow,K}=\pm 1/2$ and $C_{\uparrow/\downarrow,K'}=\pm 1/2$. Accordingly, the changes of valley-projected spin Chern numbers across the domain wall are $\Delta C_{\uparrow/\downarrow,K}=\mp 1$ and $\Delta C_{\uparrow/\downarrow,K'}=\mp 1$, which implies from the bulk-edge correspondence that there will be one spin-up edge state with negative group velocity and one spin-down edge state with positive group velocity in each of K and K' valleys. This theoretical analysis is confirmed by the simulated band diagram in Fig. 4a; each spin state occupies both valleys, and the group velocity of topological edge states is determined by spin, not valley.

We excited the domain wall with a dipole antenna at the left port (marked as a star in Fig. 4b). Using a loop antenna mounted on a robotic arm to scan the spatial distribution of $H_z$ field, we captured two out-coupled beams inside the empty waveguide region as shown in Fig. 4b. Applying phase-matching condition (see details in Supplementary Information), one can understand that the two out-coupled beams correspond to the two valleys, respectively. Then we measured the phase difference between $E_z$ and $H_z$ along the domain wall, whose results stays in the vicinity of $\pi$ as shown in Fig. 4c. This shows that the right-moving domain-wall edge states are spin-down polarized. Exchanging the excitation and detection positions, the phase difference between $E_z$ and $H_z$ now switches to close to 0. In other words, the switch of propagating directions from right-moving to left-moving converts the edge states to be spin-up polarized.

We then take $(\theta, g) = (30°, 1.1 \text{ mm})$ (denoted as point B in Fig. 2b) to demonstrate the QVH phase. The domain wall is shown in the inset of Fig. 4d. In this case, the valley-projected spin Chern numbers in the upper domain are $C_{\uparrow/\downarrow,K}=1/2$ and $C_{\uparrow/\downarrow,K'}=-1/2$, and in the lower domain, $C_{\uparrow/\downarrow,K}=-1/2$ and $C_{\uparrow/\downarrow,K'}=1/2$. Since now the changes of valley-projected spin Chern numbers across the domain wall are $\Delta C_{\uparrow/\downarrow,K}=1$ and $\Delta C_{\uparrow/\downarrow,K'}=-1$, the K (K') valley shall host both spin states propagating with positive (negative) group velocities; in other words, the group velocity of topological edge states is determined by valley, not spin.

In contrast to the QSH phase, the captured out-coupling TE waveguide mode from a QVH domain wall only shows a single beam as in Fig. 4e. This means the right-moving edge states are locked to only one valley. Moreover, the measured phase difference between $E_z$ and $H_z$ now stays between 0 and $\pi$ as shown in Fig. 4f; this phase difference will also vary along the domain wall (shown in Supplementary Information). The is because two nondegerante spins are excited for the edge states. Therefore the phase difference between $E_z$ and $H_z$ is not locked to any specific value.

The above results demonstrate a new class of PTIs, in which not only spin and valley are controllable by each other, but the topological phase can also be altered from a QSH phase to a QVH phase. The competition between spin- and valley- induced gauge fields can merge spintronics and valleytronics in the same photonic platform, and result in novel phenomena such as spin-valley–coupled Klein tunnelling[19]. The ability to tune the topology with lattice parameters offers more flexible ways to construct reconfigurable photonic topological phases, and can also be exploited to study new phenomena that are difficult in condensed matter systems such as programmable topological Moiré patterns[33]. Furthermore, as suggested in the silicene system[28], introducing time-reversal-symmetry breaking term (which can be achieved by using gyromagnetic materials) can lead to even richer topological phases.

**Methods**
**Simulation**. The band structures are simulated with first-principle simulation software COMSOL Multiphysics, RF module. The aluminum tripods as well as the upper and lower metal plates used in experiments are modelled as perfectly electric conductor (PEC). All other boundaries are set to satisfy Floquet boundary condition. In edge bands simulations, there are 75 tripods for each domain and the eigenstates related to the edge states at the upper and lower interfaces are removed. The field patterns are simulated with CST Microwave Studio.
**Fabrication.** The tripods made of aluminum are fabricated with wire Electric Discharge Machining (wire EDM) method. The air gaps are filled with a foam spacer (thickness 1.1 mm, ROHACELL 71 HF)).

**Acknowledgements**
This work was supported by the Young Thousand Talent Plan, China, National Natural Science Foundation of China under Grants No. 61801426, and Singapore Ministry of Education under Grant No. MOE2015-T2-1-070, MOE2015-T2-2-008, MOE2016-T3-1-006, and Tier 1 RG174/16 (S).

**Author Contributions**
F. G., H. X., G. S. and B. Z. conceived the idea. H.X., F. G., and Y. Y. performed simulations. F. G. did the theoretical explanation. H. X. performed the experiment. F. G., Y. C., G. S., and B. Z. supervised the project.

# References

1. Lu, L., Joannopoulos, J. D. & Soljačić, M. Topological photonics. *Nat. Photonics* **8**, 821–829 (2014).

2. Khanikaev, A. B & Shvets, G. Two-dimensional topological photonics. *Nat. Photonics* **11**, 763–773 (2017).

3. Ozawa, T. *et al*. Topological photonics. Preprint at https://arxiv.org/abs/1802.04173 (2018).

4. Khanikaev, A. B. *et al.* Photonic topological insulators. *Nature Mater*. **12**, 233-239 (2012).

5. Chen, W. *et al.* Experimental realization of photonic topological insulator in a uniaxial metacrystal waveguide. *Nature Commun.* **5**, 5782 (2014).

6. Ma, T., Khanikaev, A. B., Mousavi, S. H. & Shvets, G. Guiding electromagnetic waves around sharp corners: topologically protected photonic transport in metawaveguides. *Phys. Rev. Lett.* **114**, 127401 (2015).

7. Cheng, X. *et al*. Robust reconfigurable electromagnetic pathways within a photonic topological insulator. *Nat. Mater.* **15**, 542-548 (2016).

8. Wu, L.-H. & Hu, X. Scheme for achieving a topological photonic crystal by using dielectric material. *Phys. Rev. Lett*. **114**, 223901 (2015).

9. Yves, S. *et al.* Crystalline metamaterials for topological properties at subwavelength scales. *Nat. Commun*. **8**, 16023 (2017).

10. Yang, Y. *et al*. Visualization of a Unidirectional Electromagnetic Waveguide Using Topological Photonic Crystals Made of Dielectric Materials. *Phys. Rev. Lett*.

**120**, 217401 (2018).

11. Barik, S. *et al*. A topological quantum optics interface. *Science*. **359**, 666–668 (2018).

12. Ma, T., & Shvets, G. All-Si valley-Hall photonic topological insulator. *New J. Phys.* **18**, 025012 (2016).

13. Ma, T., & Shvets, G., Scattering-free optical edge states between heterogeneous photonic topological insulators. *Phys. Rev. B* **95**, 165102 (2017).

14. Gao, F. *et al*. Topologically protected refraction of robust kink states in valley photonic crystals. *Nat. Phys.* **14**, 140-144 (2018).

15. Wu, X., *et al*. Direct observation of valley-polarized topological edge states in designer surface plasmon crystals. *Nature Commun.* **8**, 1304(2017).

16. Gao, Z. *et al*. Valley surface-wave photonic crystal and its bulk/edge transport, *Phys. Rev. B*. **96**, 201402 (2017).

17. Noh, J., Huang, S., Chen, K. P. & Rechtsman, M. C. Observation of Photonic Topological Valley Hall Edge States. *Phys. Rev. Lett.* **120**, 063902 (2018).

18. Dong, J., Chen, X.-D., Zhu, H., Wang, Y. & Zhang, X. Valley photonic crystals for control of spin and topology. *Nat. Mater*. **16**, 298–302 (2016).

19. Ni, X. *et al*. Spin- and valley-polarized one-way Klein tunneling in photonic topological insulators. *Sci. Adv*. **4**, eaap8802 (2018).

20. Kane, C. L., Mele, E. J., Quantum Spin Hall Effect in Graphene. *Phys. Rev. Lett.* **95**, 226801 (2005).

21. Bernevig, B. A., Hughes, T. L., and Zhang S., Quantum Spin Hall Effect and

Topological Phase Transition in HgTe Quantum Wells. *Science* **314**, 1757-1761 (2006).

22. Zhou, S. *et al*. Substrate-induced bandgap opening in epitaxial graphene. *Nature Mater.* **6**, 770-775 (2007)

23. Xiao, D., Yao,W. & Niu, Q. Valley-contrasting physics in graphene: magnetic moment and topological transport. *Phys. Rev. Lett.* **99**, 236809 (2007).

24. Ju, L. *et al.* Topological valley transport at bilayer graphene domain walls. *Nature* **520**, 650-655 (2015).

25. Li, J. et al. Gate-controlled topological conducting channels in bilayer graphene. *Nat. Nanotech*. **11**, 1060-1065 (2016).

26. Schaibley, J. R. *et al*. Valleytronics in 2D materials. *Nat. Rev. Mater*. **1**, 16055 (2016).

27. Manzeli, S. *et al*. 2D transition metal dichalcogenides. *Nat. Rev. Mater*. **2**, 17033 (2017).

28. Ezawa, M. Photoinduced Topological Phase Transition and a Single Dirac-Cone State in Silicene. *Phys. Rev. Lett.* **110**, 026603 (2013).

29. Slobozhanyuk, A. P. *et al.* Experimental demonstration of topological effects in bianisotropic metamaterials. *Sci. Rep.* **6**, 22270 (2016).

30. Deng, F. *et al.* Observation of valley-dependent beams in photonic graphene. *Opt. Express,* **22**, 23605 (2014).

31. Chen, X., Deng, W., Lu, j. & Dong, J. Valley-controlled propagation of pseudospin states in bulk metacrystal waveguides. *Phys. Rev. B* **97**, 184201 (2018).

32. Ezawa, M. Topological Kirchholff law and bulk-edge correspondance for valley Chern and spin-valley Chern numbers. *Phys. Rev. B* **88**, 161406 (2013).

33. Tong, Q. *et al*. Topological mosaics in moiré superlattices of van der Waals heterobilayers. *Nat. Phys.* **13**, 356-362 (2017).

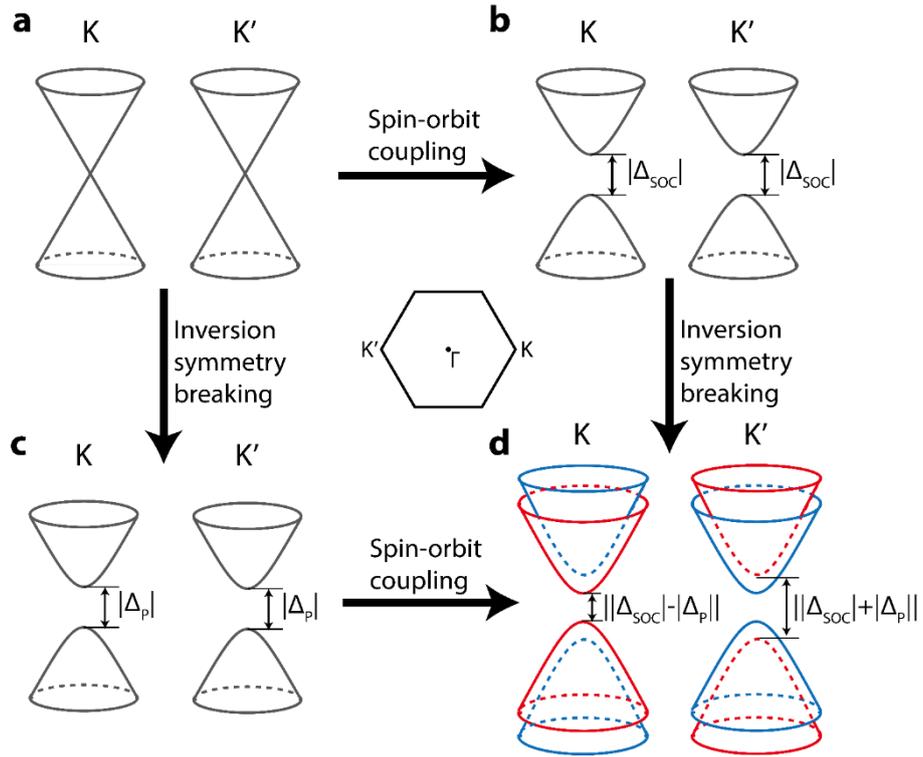

**Figure 1 | Mechanism of topological bandgap formation by different perturbations.**
**a**, Band spectrum of an unperturbed honeycomb system with Dirac cones located at K and K' valleys. **b,** When spin-orbit coupling is introduced, a topological bandgap is open, exhibiting the quantum spin Hall phase. **c**, When inversion symmetry breaking is introduced, the opened bandgap possesses topological properties similar to the quantum valley Hall phase. **d**, When both two perturbations are introduced, the degeneracy between the two spin states will be lifted and the split of two spin bands will be valley dependent. The blue and red colours stand for spin-up and spin-down states. The centre inset indicates the first Brillouin zone.

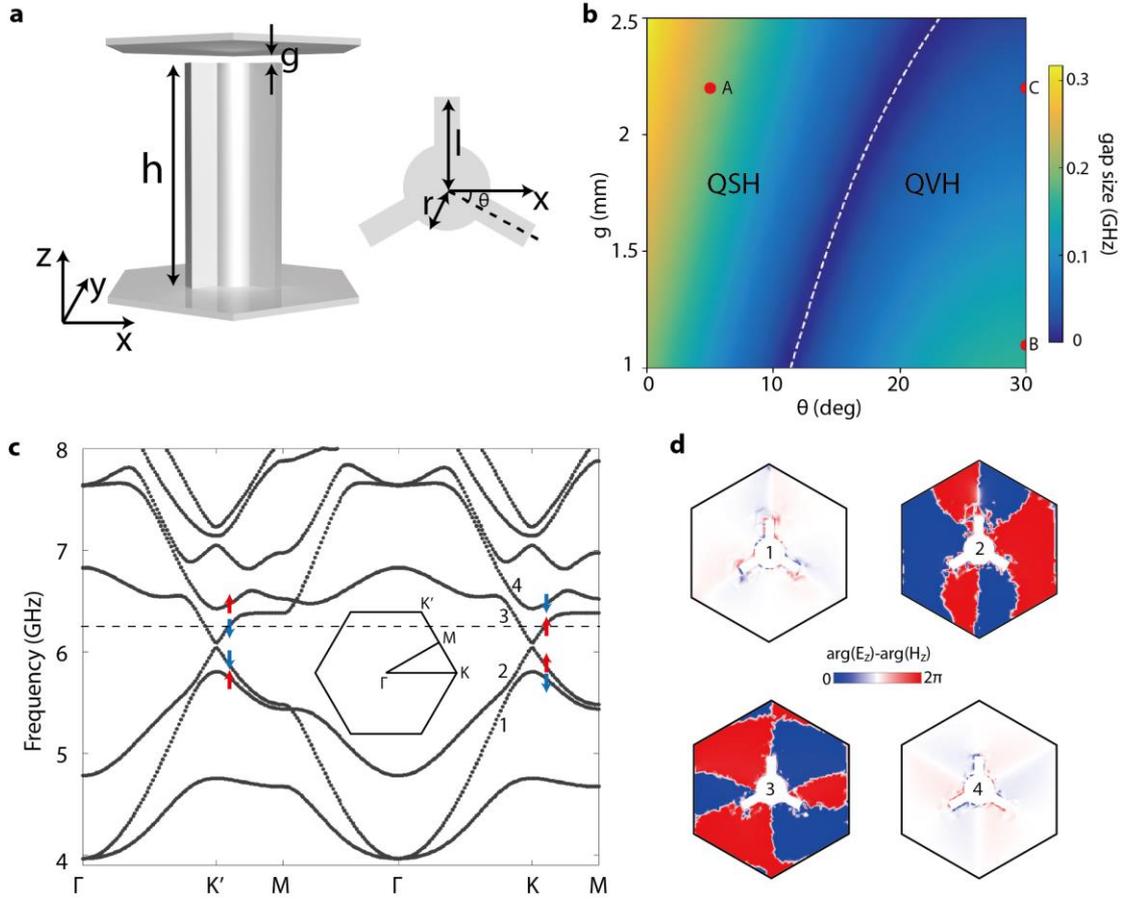

**Figure 2 | Properties of spin-valley-controlled photonic topological insulator. a**, Schematic of the unit cell, where a metallic tripod is sandwiched by two parallel metal plates. A gap *g* exists between the tripod and the top plate. The tripod is oriented with an angle *θ* with the *x* axis. Other parameters for the tripod are *h*=34.6 mm, *r*=3.68 mm and *l*=7.95mm. The lattice constant *a*=36.8 mm. **b**, Phase diagram of the photonic topological insulator in terms of *g* and *θ*. The dashed white line denotes a boundary of zero bandgap size, which separates the two topologically distinct QSH and QVH phases. **c**, Band diagram with *g*=2.2 mm and *θ*=30° (labelled as point C in **b**). Red and blue arrows indicate the spin up and spin down states, respectively. **d**, Phase difference between $E_z$ and $H_z$ fields at K point for the bands labelled as '1', '2', '3' and '4' in **c**. The plotted figures are at the middle plane of the unit cell.

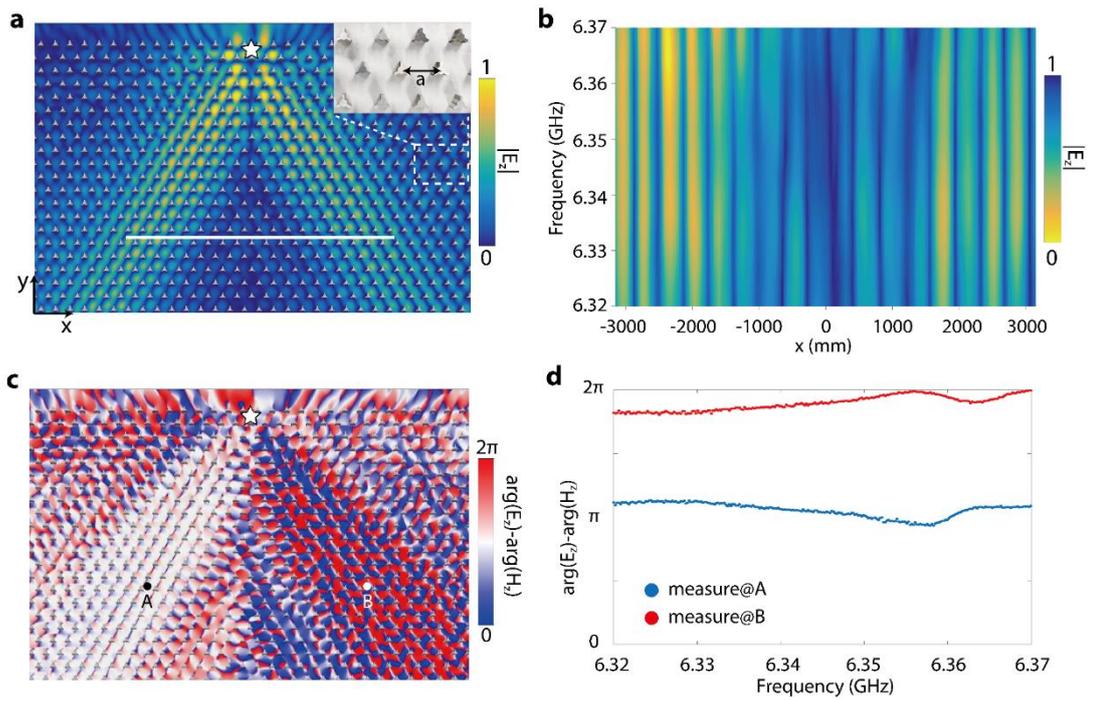

**Figure 3 | Demonstration of spin-valley locking. a**, Simulated $|E_z|$ field at 6.35GHz. Here $(g, \theta)$=(2.2 mm, 30°). The white star on the middle top of the lattice indicates the dipole source. The inset shows the photo of the lattice in measurement. **b**, Measured $|E_z|$ field distribution along the while solid line in **a**. **c**, Simulated phase difference between $E_z$ and $H_z$ fields at 6.35GHz. **d**, Measured phase difference between $E_z$ and $H_z$ fields at A and B points in **c**.

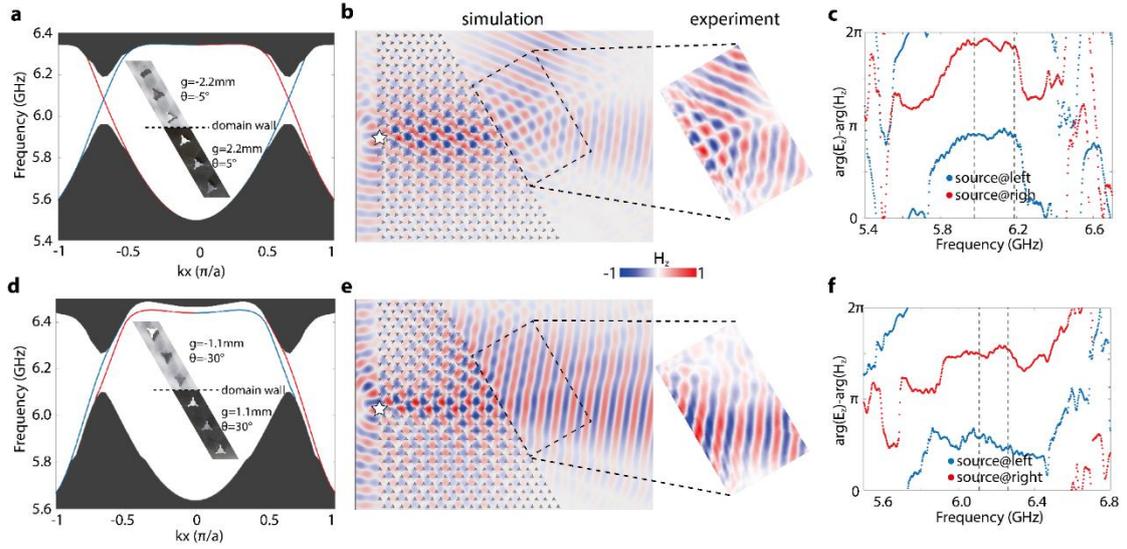

**Figure 4 | Demonstration of topological transition between QSH and QVH phases. a,** Simulated band diagram of a domain wall between two QSH-phase domains with opposite settings. The inset is a photo of the domain wall. Here the positive (negative) sign of *g* means the tripods touch the bottom (top) plate. Similarly, the positive (negative) sign of *θ* denotes the right leg of tripods are below (above) the *x* axis. The red and blue lines represent spin up and spin down for the edge states, respectively. **b,** Outcoupling of the edge states in the QSH phase through a zigzag termination at 6.08 GHz. The parameters and configurations of the domain wall are the same as in **a**. The star denotes the location of a dipole source. **c,** Measured phase difference between $E_z$ and $H_z$ fields at the end of the domain wall. **d,** Simulated band diagram of a domain wall between two QVH-phase domains with opposite settings. The inset is a photo of the domain wall. **e,** Outcoupling of the edge states in the QVH phase through a zigzag termination at 6.18 GHz. The parameters and configurations of the domain wall are the same as in **d**. The star denotes the location of a dipole source. **f,** Measured phase difference between $E_z$ and $H_z$ fields at the end of the domain wall.

17